# RAS Specialist Discussion Meeting report

**David Kuridze**, **Lyndsay Fletcher** and **Hugh Hudson** report on the RAS Specialist Discussion Meeting '3D Structure of the Flare Chromosphere'.

Solar flares and associated coronal mass ejections are the largest explosions in the solar system, releasing enormous amounts of energy, ~$10^{32}$ ergs in just a few minutes, equivalent to 25000 times the Earth's current total annual energy consumption (2019 value = 418 exajoules) and comparable to the energy of the Chicxulub asteroid impact (implicated in serious dinosaur loss). Flares are central to solar system physics since they drive space weather. They are important for understanding other star systems and of great interest to fundamental plasma physics.

Flares originate in magnetic active regions, and energy is released through the reconfiguration of the coronal magnetic field, involving one of the most fundamental processes in astrophysical plasmas: magnetic reconnection. The sudden energy release heats solar plasmas to up to at least 20MK and creates the flare arcade, a system of magnetic loops with lower energy configuration, along which a significant amount of the released energy is transported via accelerated particles, thermal conduction and/or MHD waves to the lower layers of the solar atmosphere (Hudson 2011). Most of this energy is dissipated on reaching the dense footpoints of the magnetic loops, giving rise to the most important observational signature of solar flares, the striking intensity enhancements/brightenings in the lower solar atmosphere (figure 1), either as small 'kernels' or elongated 'ribbons'.

Photospheric kernels were discovered and reported first by British astronomers Richard Carrington and Richard Hodgson (Carrington 1859; Hodgson 1859) during one of the most famous astronomical events – a sudden dramatic brightening at the solar photosphere recorded on 1 September 1859 within a large sunspot group. Later observations, including space-borne telescopes, have shown that flare sources are present through all layers of the solar atmosphere and emit broadly throughout the whole electromagnetic spectrum from long radio wavelengths to GeV gamma-rays (Fletcher *et al.* 2011). However, it is accepted that the vast majority of the radiative energy of flares originates from the chromosphere, making them key features to understand in the physics of solar flares, and vital diagnostics for the processes of energy dissipation.

There have been tremendous observational and numerical modelling advances over recent decades, but many aspects of the flare chromosphere, such as its response to heating, radiation mechanisms in different spectral domains, three-dimensional (3D) geometry and fine-scale structure, remain the subject of ongoing debate and research. Spectroscopy of flare ribbons offers the most powerful opportunity to investigate and diagnose the flare chromosphere. However, modelling and interpretation of chromospheric flare emission are extremely challenging, as it involves solving the radiative transfer problem in conditions where local thermodynamic equilibrium does not hold. The impulsive deposition of huge amounts of energy localized in space and time also means that the flare chromosphere is extremely dynamic, exhibiting intense upflows (hot, explosive expansion, misleadingly called 'evaporation'), downflows (cool condensation) and shocks redistributing

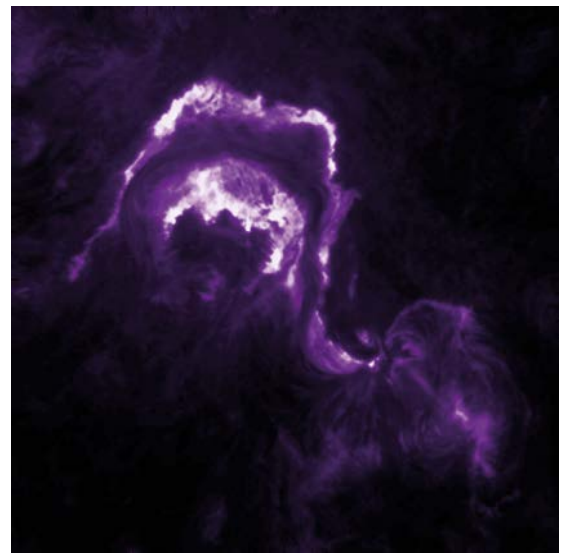

**1** *Flare ribbons detected in the chromosphere in the 304 Å filtergram.* (NASA/SDO)

the mass and energy in the lower atmosphere. Accurate interpretation and diagnostics of the flare chromosphere requires radiative hydrodynamic (RHD) or radiative magnetohydrodynamic (RMHD) computations (solving HD or MHD equations combined with radiative transport (RT) equations) where all these effects intersect.

The aim of the RAS Specialist Discussion Meeting '3D Structure of the Flare Chromosphere' was to discuss flare chromospheric observations, state-of-the-art simulations, recent advances, and challenges as well as new observing opportunities offered by recently commissioned and future solar telescopes.

### Flare spectroscopy

The meeting began with an invited review given by **Mihalis Mathioudakis** (Queen's University Belfast), on the diagnostic potential of asymmetric flare line profiles.

The intense dynamic processes (evaporation/condensation) resulting from strong heating lead to spectral line asymmetries, which can be detected directly in spectroscopy (Gunn *et al.* 1994; Berlicki 2007) and used to investigate evolution of the vertical structure of the flare velocity field through the atmosphere. Comparisons between observed and synthetic spectra calculated with RHD models suggest that a multicomponent velocity field in the chromosphere creates substantial differences in the opacities between the red and blue wings of chromospheric lines, including the hydrogen Balmer lines. As a result, asymmetric chromospheric spectral lines are produced (Ichimoto & Kurokawa 1984; Heinzel *et al.* 1994; Kuridze *et al.* 2015).

Turning to spectropolarimetry, Mihalis discussed how measurements of the all-important coronal magnetic field can be made in flare loops at the solar limb. Heated chromospheric plasma is first 'evaporated' into coronal loops, and as it cools it radiates strongly in optical and near-infrared spectral regions. These can be observed in detail with modern high-resolution ground-based telescopes (figure 2), and spectropolarimetric techniques applied. Kuridze *et al.* (2019) succeeded in observing and mapping the magnetic field strength with unprecedented accuracy and resolution in coronal loops formed in the aftermath of a large solar flare (figure 2). On the solar disk the situation is more difficult, since the formation height of the radiation cannot be seen directly. Mihalis pointed out that apparent variations of magnetic field on disk could be due to flare-induced changes in the line formation height revealing the field strength at a different location, rather than changes of magnetic field at a fixed location (Hong *et al.* 2018; Kuridze *et al.* 2018; Nelson *et al.* 2021). Finally, it was mentioned that the effect of stellar flares on spectral lines can be a source of 'jitter' in the measured radial velocity, interfering with exoplanet detection (Saar *et al.* 2018). This can be studied by looking in detail at the formation of solar photospheric spectral lines.

This was further explored by **Aaron Monson** (Queen's University Belfast) looking at the deep-forming photospheric spectral lines in solar and stellar flares.

RHD simulations show that flare-induced velocities create asymmetries in photospheric spectral lines, e.g. in Fe I 6301 Å (Monson *et al.* 2021). In the non-flare atmosphere, this line forms in the photosphere. However, during the flare, the upflowing chromosphere can contribute significantly to the line formation; absorbing upflowing material creates 'fake' redshifts in the line profile. Aaron also showed that in stellar flare models Fe I spectral lines can provide very useful diagnostics for downwardly moving chromospheric condensations.

The flare chromosphere has non-uniform, fine-scale horizontal structures, but due to the requirements of modelling (non-LTE radiative transfer, non-equilibrium ionization, hydrodynamic shocks etc.) current flare simulations are normally 1D, sacrificing the higher spatial dimensions and effect of horizontal radiative transfer.

**Chris Osborne** (University of Glasgow) discussed the importance of moving beyond this, presenting time-dependent two-dimensional radiative models of a slab of quiet Sun atmosphere irradiated by an adjacent flare volume.

It was shown that irradiation of the non-flaring chromosphere by the flare volume can significantly affect the H$\alpha$ and Ca II 8542 Å lines, with effects persisting at distances up to 1000 km from the flare source. However, the continuum emission forming in the deeper layers of the solar atmosphere is unchanged. These effects will be important in estimates of the filling factor of emitting plasma and energetics of the flare chromosphere, as detected intensity enhancements could be 'reprocessed' radiation, and not associated with the direct flare heating.

**Shaun McLaughlin** (Queen's University Belfast) discussed the formation of the Lyman continuum in the flare chromosphere.

In contrast to the quiet Sun, where this line forms in the transition region (above the chromosphere), RHD flare models show that it forms deeper in the chromosphere under LTE conditions, making it a powerful tool for probing the chromospheric flare. The SPICE instrument on board the recently launched Solar Orbiter mission will be able to image the Lyman continuum emission, opening new, important avenues for future flare research.

The vertical structure of the chromosphere can be probed by emission formed at different heights, shedding light on the long-standing problem of how flare energy is dissipated in the chromosphere. Different energy transport and dissipation mechanisms predict different heights of energy input. For example, conductive energy input affects mostly the upper atmosphere. Electron beams are stopped by Coulomb collisions in the chromospheric collisional 'thick target', and their penetration depends on their typical energy, while Alfvén waves might dissipate by ion-neutral friction in the mid-chromosphere. Limb observations of the flare chromosphere offer a unique opportunity to investigate the formation height of the chromospheric ribbon.

**Paulo Simões** (Mackenzie Presbyterian University) presented limb observations using the 1600 Å and 1700 Å ultraviolet (UV) channels of the Solar Dynamics Observatory Atmospheric Imaging Assembly (SDO/AIA).

Using broad-coverage spectral data from Skylab, Paulo identified C IV and C I line emissions as the main contributors to the flare emissions in the wide-band AIA 1600 and 1700 Å channels, respectively. These lines form at chromospheric temperatures so the 1600 and 1700 Å channels can be used for the diagnostics of chromospheric ribbons. The observed vertical profile of the ribbons at the limb indicates a formation height of the UV ribbon emission of about 1 Mm above the photosphere. Similar results have been reported by Kuridze *et al.* (2020) with high-resolution SST observations of off-limb flare in the H$\beta$ line (figure 2).

Multi-wavelength spectroscopy is an effective diagnostic tool for distinguishing between different models for flare energy transport, as the location and strength of the energy input plays a significant role in shaping the observed spectral line profiles.

**Alberto Sainz Dalda** (Bay Area/Lockheed-Martin Solar and Astrophysics Laboratory) showed that the multi-spectral capabilities of Interface Region Imagaing Spectrograph (IRIS) lines can provide useful constraints on atmospheric models.

They can recover thermal and non-thermal velocities (from spectral line broadening, e.g. Kerr *et al.* 2015; Rubio da Costa *et al.* 2016; Jeffrey *et al.* 2018) as well as the temperature and electron density of the flare

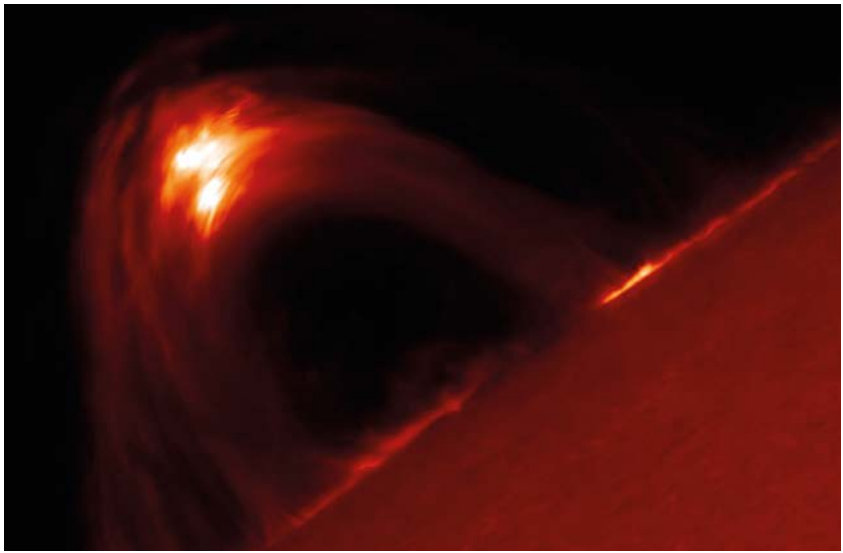

**2** *The flare coronal arcade captured by the Swedish Solar Telescope (La Palma, Spain) at 0.45 Å from the centre of the H$\beta$ line core during the famous 10 September 2017 X8.2 class flare. The morphology of the flare event suggests that the energy propagates down along the flare arcade legs from the apex and heats the chromosphere near the footpoints. Therefore, brightenings near footpoints are interpreted as the off-limb flare ribbons projected on the limb and seen from the side.*
(adapted from Kuridze et al. 2020)

atmosphere. Interpretation of the lines requires advanced methods including fitting of multiple spectra with a model atmosphere using state-of-the-art inversion algorithms. Alberto demonstrated that inversion of the C II and Mg II h & k flare line profiles can be used to derive non-thermal velocities associated with turbulence of the flare chromosphere. Machine-learning on IRIS flare data was used to identify spectral profiles showing a single peak in Mg II h & k plus Mg II UV triplet lines. These are associated with preflare activity, with the increased temperature and electron density in the chromosphere governing their properties (Woods *et al.* 2021).

IRIS multi-wavelength spectroscopic data was also discussed by **Juraj Lorincik** (Bay Area/Lockheed-Martin Solar and Astrophysics Laboratory), who presented data on line shifts in cool chromospheric lines.

Redshifts in these lines due to the condensation downflows produced by the flare overpressure (figure 3) are well known (Graham & Cauzzi 2015), but in the event presented, fast-moving flare kernels appearing at the beginning of the flare show both redshifts and blueshifts in the Si IV, Mg II and C II lines. The blueshifts are extremely short-lived, with 'normal' redshifted behaviour being restored quickly. Flare models suggest that the short-lived blueshifts are due to heating of a deep chromospheric layer, pushing upwards the chromosphere that emits in these lines. The heated layer also pushes down the chromosphere underneath and produces observed redshifts that become dominant very quickly after the flare (Polito *et al.* 2018; Tei *et al.* 2018).

**Adam Kowalski** (National Solar Observatory/University of Colorado) gave the second invited talk on the spectral signatures of the 3D flaring chromosphere, with an emphasis on high-time resolution model predictions of hydrogen emission line broadening.

Adam started his talk with the review of the 'time-scale' problem. Chromospheric condensations (fast downflows) produced as a result of explosive heating appear in spectra as red-wing asymmetries in hydrogen Balmer lines. Models estimate the half-life of the condensation (time that it takes for the maximum downward velocity of the condensation to drop by half) as 4-8 seconds, much less than is observed (Fisher 1989). This inconsistency could be related to the spatial resolution of the observations, if multiple, non-simultaneous, condensation sources exist in a single pixel, and improved observations from new telescopes are needed to further examine this.

IRIS's high spectral resolution has also allowed line asymmetries to be examined. In the Mg II lines a red-wing asymmetry appears first as a satellite component that over time moves in wavelength towards the stationary component, and eventually merges (Graham *et al.* 2020). The two components have been successfully modelled with a 1D radiation hydrodynamics code (Kowalski *et al.* 2017), showing that pressure gradients in the chromosphere drive upflows and downflows over a very high density but narrow region (~25 km) (figure 3). The downflows cool rapidly from 80 000 K to 10 000 K, and when cooling they emit in the red wing of chromospheric lines such as Fe II, Mg II, and hydrogen, thereby explaining the asymmetries.

Observed line broadening can be much larger than theory and model predictions. One possible explanation is collisional broadening, in particular Stark broadening (energy level splitting linearly proportional to the microscopic electric field generated by ambient charges, both electrons and ions). The

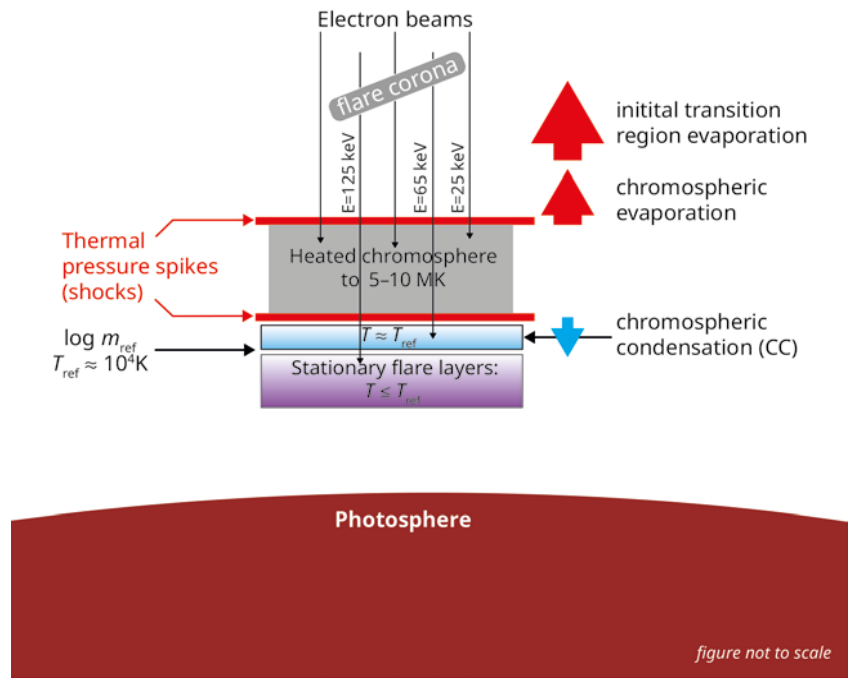

**3** *Schematic picture illustrating basic features that develop in the flare atmosphere as the response of electron beam heating (thin black arrows).*
(from Kowalski & Allred 2018)

IRIS Mg II h & k lines in some flare locations and times have the Lorentzian shape expected of the collisional mechanism. However, even when 1D simulations are run with quadratic Stark broadening the Stark widths for Mg II h & k are still some 30 times smaller than observed (Zhu *et al.* 2019). Hydrogen emission lines in flares are also broadened dramatically due to the Stark effect, with broadening increasing in higher-order lines such as Balmer gamma and delta (Kowalski *et al.* 2022). Comparison between hydrogen line models and observations of solar/stellar flare lines suggest that an updated treatment, developed originally for white dwarf flare models (Tremblay & Bergeron 2009; Kowalski *et al.* 2022) is needed. 'Macroturbulent' broadening is also a possibility, in which the superposition of several different sources within one observed resolution element could result in broadening, due to the superposition of the different velocity fields in these sources. This implies a need for multi-dimensional or multi-thread modelling approaches (Warren 2006).

The new four-metre Daniel K Inouye Solar Telescope (DKIST) offers the possibility of resolving some of these problems. Figure 4 demonstrates the huge improvements in spatial resolution showing how Hurricane Teddy would be resolved hypothetically from the Sun with IRIS and with DKIST. The unprecedented high spatial, temporal and spectral resolution of DKIST and broader wavelength coverage for the chromospheric spectral lines in the optical will allow detailed investigation of flare macrophysics and the line-broadening mechanisms. Furthermore, high-resolution solar flare data of the hydrogen lines will provide extremely important constraints on models of spatially unresolved stellar flare spectra.

**Hugh Hudson** in his poster presentation discussed the prospects for the use of the mm/IR free-free continuum for flare chromosphere diagnostics.

The opacity of far-infrared to microwave and radio regions is mostly due to the free-free continuum (Simões *et al.* 2017). This opacity grows steadily toward the near infrared (1.6 microns) into the microwave/radio, implying that flare sources in the lower atmosphere might not be observable in microwaves. However, modelling short timescales indicates that one can observe the altitudes of electron beam penetration in the mm region, for example at 100 GHz. The analyses show that the main source of the emission (layer where

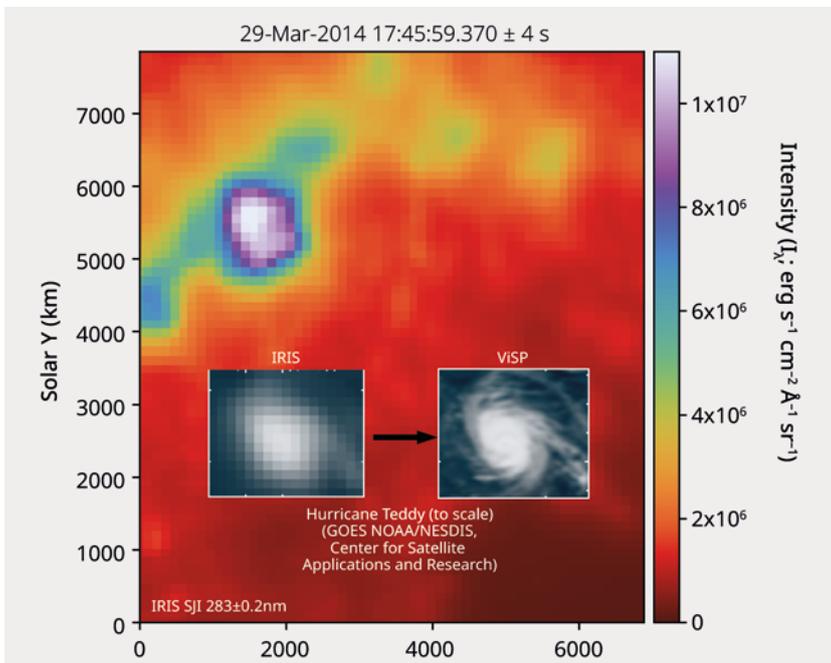

**4** *Flare ribbons observed with the IRIS satellite. Grey images show how Hurricane Teddy would have been observed with IRIS and DKIST/ViSP spectrograph if we put them on the Sun.* (Adam Kowalski) optical depth is close to 1) of free-free continuum is located deep in the solar atmosphere. The beam heating model creates thermal signatures, bright chromospheric footpoints in the long wavelength spectral range, suggesting that flare ribbons can be detected and analyzed in the microwave/radio spectrum. Such observations can be done with ALMA, the highest-resolution mm/submm telescope.

In the next contributed talk, **Petr Heinzel** (Astronomical Institute of the Czech Academy of Sciences) discussed emission in dense flare coronal loops.

Some high-density flare loops can be detected in the visible continuum, and analysis of white-light data from the X9.3 solar flare in SOL2017–09–10 from the *SDO* Helioseismic and Magnetic Imager showed that the density of the flare loops are around $10^{12-13}$ cm$^{-3}$ (Jejčič *et al.* 2018) which is almost two orders of magnitude higher than typical coronal-loop densities (Young *et al.* 2009). In cool extreme UV (EUV) images, flare loops can appear dark, traditionally interpreted as absorption of background EUV radiation. Analysis of the September 2017 off-limb flare loops using different *SDO*/AIA channels and Swedish Solar Telescope data in Hβ and ionized calcium indicates that the helium recombination continua also contribute to cool flare loop emission (Heinzel *et al.* 2020).

The contribution of flare loops to the white light (WL) emission of stellar super flares, reported intensively during recent decades, was discussed by **Julius Koza** (Astronomical Institute, Slovak Academy of Sciences, Tatranská Lomnica, Slovakia) in his contributed talk.

The thermodynamic parameters of the SOL2017-09-10 flare loops were obtained from the SST spectral imaging data in the Ca II 8542 Å and the Hβ lines and the non-LTE spectral inversions, confirming a high electron density of observed flare loops (Koza *et al.* 2019). The presence of such high densities in solar flares supports the scenario in which WL emission of flare loops could make significant contributions to the emissions of WL optical continuum of stellar superflares, as suggested by Heinzel & Shibata (2018).

**Sargam Mulay** (University of Glasgow) presented observations of flare ribbons in molecular hydrogen lines. The H$_2$ emission formed in the coolest parts of the flare atmosphere could be fluorescently excited by Si IV 1402.77 Å UV radiation. A strong spatial and temporal correlation between H$_2$ and Si IV emission has been found indeed during the impulsive phase of a flare observed with the IRIS satellite (Mulay & Fletcher 2021). H$_2$ emission provides a unique diagnostic of the temperature minimum region of the flare atmosphere. The temperature of this region (~4500K) is too low to populate upper hydrogen levels that makes it inaccessible for atomic hydrogen lines.

### Flare studies with machine learning

Due to the huge amount of flare data with multi-instrument, multiwavelength spectropolarimetric observations covering ribbons, machine learning techniques are becoming more and more popular in modern solar flare physics.

**Brandon Panos** (The University of Geneva) presented an analysis of k-means clustering applied to several million IRIS spectra from 21 large M- and X-class flares.

This revealed characteristic spatial and temporal patterns (Panos & Kleint 2020). The leading edges of all flare ribbons have similar centrally reversed Mg II line profiles (Panos *et al.* 2021) and similar patterns of time evolution going from: (1) broad, single peaked; (2) blue-shifted and centrally reversed profiles during the impulsive phase of the flare; (3) relaxed single-peaked profiles; and, (4) strongly red-shifted profiles corresponding to cool downflows of post-flare plasma. The analysis also reveals that the blue-shifted central reversal appears simultaneously in the C II and Si IV lines.

**Jonas Zbinden** (University of Geneva) gave a poster presentation describing machine learning for solar flare prediction.

Neural networks can be trained to attempt the prediction of solar flares but a problem with flare data-sets is that they always contain non-flaring pixels, which may negatively affect any prediction models. Jonas used a variational autoencoder to pre-process solar spectra and clean observations of quiet Sun spectra which have no predictive power for solar flares. In future this will help to identify the spectral signatures the neural network needs to focus on for flare predictions.

An alternative approach for solar flare prediction was discussed in poster presentations by **Marianna Korsos** (Aberystwyth University) and **Aabha Monga** (Aryabhatta Research Institute), using methods based on the measurements of topological properties of magnetic field such as magnetic helicity and twist in the flaring active regions.

It has been proposed that the evolution of magnetic helicity flux and its oscillatory behaviour at different atmospheric heights can be used to identify its role in the dynamics of active regions and investigate if they can provide valuable clues for flare prediction.

### Flare morphology and topology

**Jaroslav Dudík** (Astronomical Institute of the Czech Academy of Sciences) discussed the saddle-shaped appearance of flare loop arcades.

He presented satellite images of five flares formed in different magnetic environments and with different energy classes (including the previously mentioned SOL2017–09 flare) all showing saddle-shaped arcades. The shape is due to the presence of relatively longer loops at the ends of the arcade (Lörinčík *et al.* 2021). Stereoscopic imaging using data from SDO/AIA and STEREO-B satellites shows that such flares have extended, hook-shaped ribbons. According to 3D MHD models the formation of observed saddle-shape and evolution of ribbons can be explained by the way magnetic reconnection occurs between the erupting flux rope and the overlying coronal arcade (Aulanier & Dudík 2019).

The final invited talk, from **Jiong Qiu** (Montana State University), was on the fine scale structure and dynamics of flare ribbons.

She included an excellent formulation of the importance of flare ribbons in the introduction which we quote here: "Flare ribbon observations provide diagnostics of flare heating: when, where, by how much, for how long, and through what mechanism, is the flare atmosphere heated."

It is well known that the observations in hard and soft X-rays (HXR/SXR) originating from the flare footpoints are extremely valuable as they can be used to investigate flare energetics including energy deposition rates (Qiu *et al.* 2012). In the absence of high-resolution HXR/SXR observations, UV/EUV photometry and spectroscopy of ribbons offer very important tools for the study of flare heating and associated processes. Analyses of flare ribbon topology in UV/EUV indicates that chromospheric ribbons map reconnection energy release sites. Therefore, reconnection in the corona plays a significant role in the formation and dynamics of flare ribbons. Furthermore, the leading edges of flare ribbons show non-uniform and highly structured patterns defined partly by the magnetic field structure of the photosphere (Naus *et al.* 2022). To understand exactly in what extent the coronal field contributes to the flare ribbon formations, we need to have high-resolution observations of the magnetic field at multiple heights in the solar atmosphere.

Observations show that the separation of flare ribbons generally increases in time. This separating motion is a well-known phenomenon in flare physics, related to the continuous/progressive reconnection process, where the reconnection X-point reforms at greater and greater heights. As a result, the newly 'ignited' loops have footpoints located at different positions, producing the apparent separations in time (Hori *et al.* 1997). These processes were successfully reproduced by the model of flare reconnection from the high-resolution MHD simulations shown in figure 5 (Dahlin *et al.* 2022). Jiong also discussed the phenomenon known as the 'dark ribbon' – a negative flare front (enhanced absorption) observed in the He 10830 Å line (Xu *et al.* 2016). RHD models indicate that these flare-induced dimmings of the He I 10830 Å line can be reproduced by including nonthermal collisional ionization in the calculations. However, modern models cannot reproduce the long duration of observed dark ribbons (Kerr *et al.* 2021).

**Alex Pietrow** (Stockholm University) presented a study of so-called fan-shaped active region jets generated near an X9.3 flare.

Being not fully opaque in chromospheric lines, the interpretation of their spectra requires the emission from the background ribbon to be considered. This was done with a cloud model technique, where a 1D slab is backlit by continuous emission produced by the flare. Using this model, the parameters of the fan-shaped jet, such as opacity and velocity were derived (Pietrow *et al.* 2022).

**Christoph Kuckein** (Instituto de Astrofísica de Canarias) presented analyses of an M3.2 class solar flare that was observed with the Vacuum Tower Telescope at Observatorio del Teide (Spain).

Spectropolarimetric raster scans of the four Stokes

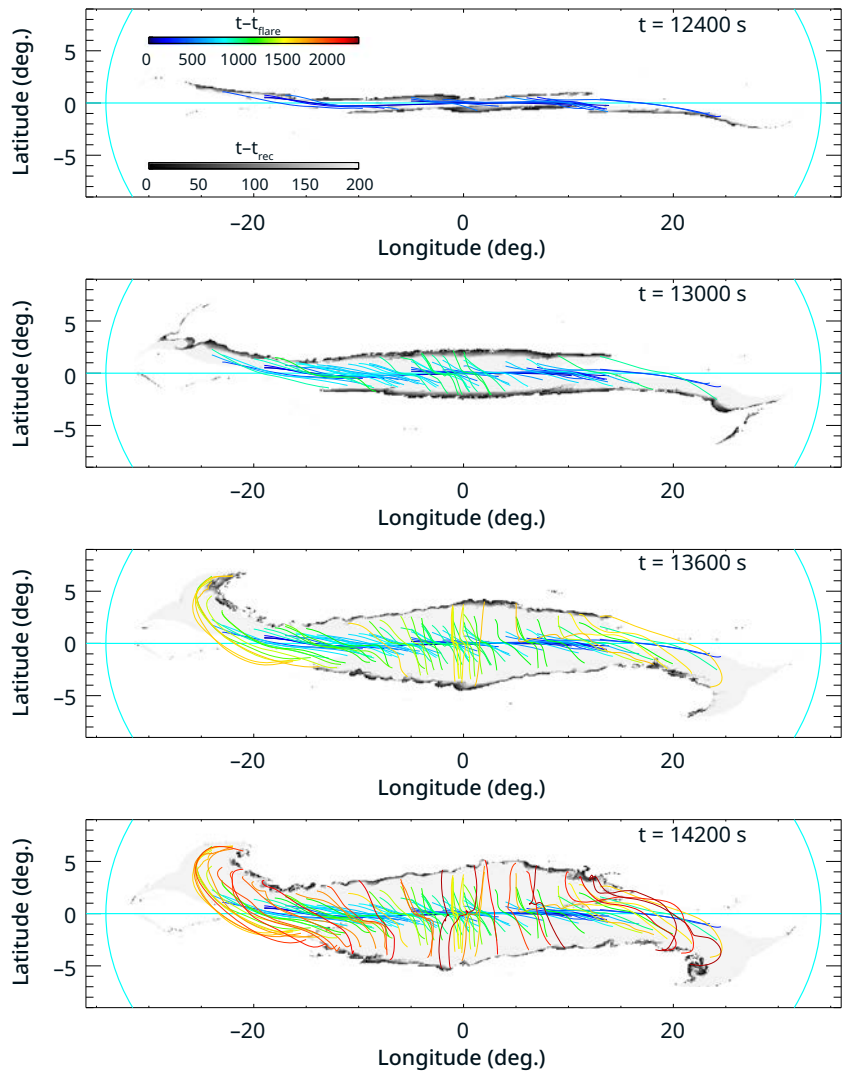

5 *MHD simulation of an eruptive flare showing evolution of the flare ribbons and loops (from Dahlin et al. 2022). Ribbons mark the footpoints of reconnected field lines (grey and color scales show the time passed after the reconnection and flare onset, $t-t_{rec}$ and $t-t_{flare}$, respectively. This explains how the separation of flare ribbons increases with time.*

parameters in the upper chromosphere (He I triplet at 1083.0 nm) were obtained with the Tenerife Infrared Polarimeter. In addition, co-temporal spectroscopic raster scans of other chromospheric spectral lines such as Ca II at 854.2 nm and Hα filtergrams were acquired with the Echelle spectrograph. Observation covered different stages of the flare including the pre-flare, impulsive, gradual, and post-flare phases. Analyses of He I line profiles revealed supersonic downflows in the impulsive phase of the flare. The complicated shape of polarization profiles suggests that overlapping components can be responsible for their formation. Furthermore, it was found that flare loops at the gradual phase are more prominent in He I and Hα than in the Ca II line.

The last contributed talk of the meeting, presented by **Tetsu Anan** (National Solar Observatory), discussed spectropolarimetric observations of flare ribbons using chromospheric lines.

The results suggest that the polarization signals and magnetic field can change during the flare due to the change in geometry (restructuring of coronal field) or formation height of magnetically sensitive chromospheric spectral lines (Kuridze *et al.* 2018; Libbrecht *et al.* 2019). Tetsu presented observations of a C4 class flare from the Domeless Solar Telescope at Hida Observatory (Japan) on 9 August 2015 in the He I triplet 1083.0 nm and Si I 1082.7 nm lines. Complicated patterns in the He I line can be reproduced with multi-component models (Anan *et al.* 2018), with analyses of the magnetic field strength of different components and their associated upward/downward velocities suggesting that two emission components are formed in the lower atmosphere where the non-thermal electron beam propagates, dissipates and heats the plasma to a coronal temperature (L1 and L2 on figure 6). An absorption component formed in the upper chromosphere is, interestingly, relatively unaffected by the beam propagation (L3 on figure 6). This deduced structure can provide important diagnostics of the flare electron beam properties.

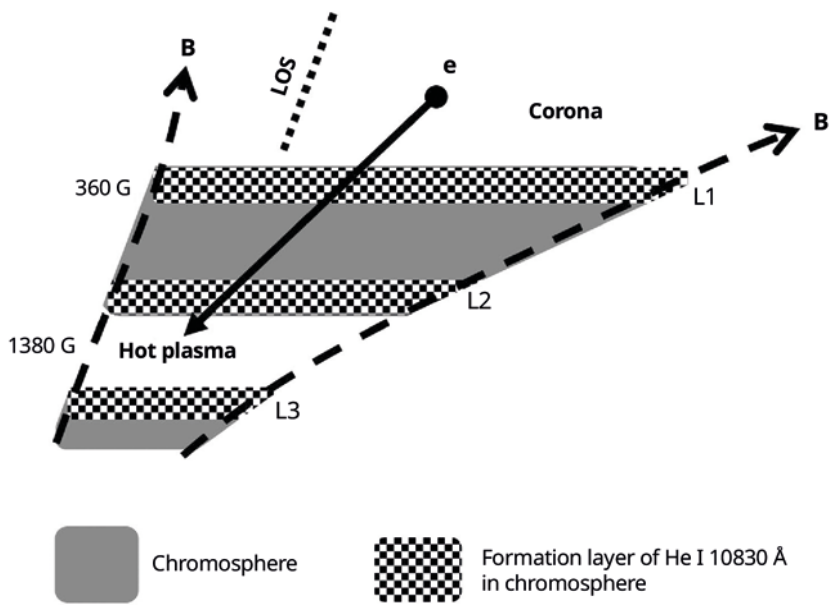

**6** *Formation heights of three components of the He I line in the flare atmosphere. The dotted line indicates the line of sight.* (from Anan et al. 2018)

## Concluding remarks

The meeting covered many important aspects of the flare chromosphere and discussed the main methods that have been used to make advances in the field during the last decades. Modern flare observations focus mostly on obtaining multi-instrument, multi-wavelength spectroscopic, spectropolarimetric and imaging data to achieve comprehensive coverage of different phenomena associated with solar flares. Combining space and ground-based observations is a very effective observational strategy, but to interpret such complex observations advanced flare models, produced with RHD/RMHD simulations are needed, and the community is developing such models. These models, assuming reasonable observational constraints, provide very realistic synthetic spectral lines of the flare chromosphere, and spectral diagnostic tools developed from them are used to interpret the observations. Atmospheric models that attempt to reproduce directly the observed flare spectral profiles in state-of-the-art RHD calculations are also a powerful way to investigate and understand flare chromospheres. Inclusion of 3D physics and improved micro-physics in numerical simulations are the next important theoretical targets. Due to the huge amount of flare observations covering ribbons, machine learning techniques are becoming an increasingly popular tool in modern flare physics as well.

Flares are a universal phenomenon detected in stars of almost all different spectral types along the main sequence. This makes this topic an excellent and unique interdisciplinary topic in astrophysics where the knowledge between solar and stellar physics communities can be easily and effectively transferred. For example, Sun-as-a-star studies have been excellent technique for the 'point-source' astronomical communities dependent on spatially unresolved observations of the stars. The meeting demonstrated that implementation of the knowledge and experience in observations and modelling of stellar flares can significantly improve observations and modelling of solar flares and vice versa.

Some of the outstanding problems of solar physics including energy transport mechanisms, physics that defines the response of the lower solar atmosphere to the energy deposition, formation of white-light emissions, magnetism and 3D fine-scale structure of the flare chromosphere will have been challenged further with improved theoretical modelling efforts and observations with recently commissioned and future high-resolution observing facilities. ●

**ACKNOWLEDGEMENTS**
The organisers would like to thank the RAS and all the participants of the meeting.

**AUTHORS**
**David Kuridze** (dak21@aber.ac.uk) is a research fellow and lecturer at Aberystwyth University, UK. His research interests include diagnostics of solar plasma, solar spectropolarimetry and magnetic fields. He has a deep affection for viewing and studying the Sun with the highest resolution telescopes.

**Lyndsay Fletcher** (Lyndsay.Fletcher@glasgow.ac.uk) is a Professor of Astrophysics at the University of Glasgow, UK, who is an enthusiast for anything to do with solar flares and their analogues.

**Hugh Hudson** (Hugh.Hudson@glasgow.ac.uk) is a general-purpose solar physicist from UC Berkeley, now happily retired in Glasgow. His current thrills involve studying the Sun as a star in various ways